\documentstyle[prd,aps]{revtex}

\input epsf

\begin{document}
\draft

\twocolumn[\hsize\textwidth\columnwidth\hsize\csname
@twocolumnfalse\endcsname

\title{Multiply Unstable Black Hole Critical Solutions}

\author{Steven L. Liebling}
\address{Center for Relativity,
         The University of Texas at Austin,
         Austin, TX 78712-1081}

\maketitle

\begin{abstract}
The gravitational collapse of a complex scalar field in
the harmonic map is modeled in spherical symmetry.
Previous work has shown that a change of stability of the attracting
critical solution occurs
in parameter space from the discretely
self-similarity critical (DSS) solution originally found by Choptuik
to the continuously self-similar (CSS)
solution found by Hirschmann and Eardley.
In the region of parameter space in which the DSS is the attractor,
a family of initial data is found which finds the CSS as its critical solution
despite the fact that it has more than one unstable mode.
An explanation of this is proposed in analogy to families that find
the DSS in the region where the CSS is the attractor.
\end{abstract}


\vskip2pc]

\section{Introduction}
\label{sec:introduction}

Much progress has been made in understanding the nonlinear
phenomena associated with the threshold of black hole formation
first found by Choptuik \cite{matt}. He studied the collapse
of a minimally coupled, real scalar field whose initial configuration
is parameterized
by some parameter $p$. For initial data with $p$ less than some
critical value $p^*$, the field energy explodes through the center of
spherical symmetry and disperses. For $p>p^*$, the configuration collapses
to form a black hole.
In the limit $p\rightarrow p^*$, a solution perfectly poised between
collapse and dispersal, the {\em critical solution}, is reached.

Naively one might expect to find a tremendous variety of critical
solutions dependent on the form of the initial data used (eg.
Gaussian pulses parameterized by an amplitude $p$, sinusoidal pulses
parameterized by a frequency $p$, etc).
However, Choptuik found that the critical solution for any interpolating
family
was precisely the same. The critical
solution he found was said to be {\em universal} because of
its apparent independence on the initial data.

In addition to its universality,
the critical solution found by Choptuik
exhibits discrete self-similarity (DSS) such that the fields
are periodic in $\log |t|$ and $\log r$ via
$f(r,t) = f(\exp(\log r+\Delta),\exp(\log |t|+\Delta))$ for a
universal constant $\Delta\approx 3.44$. Later, studies of axisymmetric vacuum
gravity revealed similar critical behavior with a different
echoing constant $\Delta \approx 0.6$, showing  $\Delta$
to be model-specific \cite{abrahams}.
However, the nature of how the field equations select a specific
value $\Delta$ was, and still is,  mysterious.

Discovery of other critical solutions has since followed including
various continuously self similar solutions (CSS) found in
perfect fluid collapse and other scalar collapse models
whereby the fields obey $f(r,t)=t^{i \omega}f_0(-r/t)$ \cite{evans,maison,eric}.

Here the issue of universality is examined.
Universality comes about
because of the presence of only one unstable (relevant) mode about the
critical solution. The single unstable mode sends nearby trajectories
in phase space away from the critical solution. The trajectories either disperse or
form black holes. This mode is then appropriately called the black hole
mode. Though the critical solution is unstable,
the process of tuning progressively
limits the influence of the mode, delaying its growth.
Modulo this unstable mode,
the critical solution is then an attractor and therefore
called an intermediate attractor.

When a critical solution has more than one unstable mode it ceases
to be an intermediate attractor. Tuning a one-parameter family of
initial data still tunes the black hole mode, but the trajectory
will still generically be sent away from such a critical
solution by the presence of the other unstable modes.

The model studied here is the harmonic map (also called the
nonlinear sigma model) with a free parameter $\kappa$.
The model represents a mapping of a self-gravitating complex scalar field
onto a
target space with constant curvature.
This curvature, parameterized by $-\kappa$,
characterizes the {\em internal} space of the field.
For the case $\kappa=0$ the model is simply a free complex scalar field
minimally coupled to gravity whose action we recover by setting $\kappa=0$
in the action for the non-linear sigma model, Eq.~(\ref{eq:action}).
In this case, the target space is the complex
plane with zero curvature.
For positive values
of $\kappa$, the non-linear sigma model maps into
a hyperboloid.
This model is equivalent to a real scalar field coupled to Brans-Dicke
gravity studied in \cite{liebling}.
For negative $\kappa$,
the target space is the sphere, $S^2$.

Because for $\kappa=0$ the model is identical to the free complex scalar
field, the attracting critical solution is known to be the DSS.
Hirschmann and Eardley
have shown
the CSS to have a pair of conjugate
unstable modes in addition to the black hole mode for $\kappa$
near zero~\cite{eric}.
However, their analysis shows that
as $\kappa$ is increased above $\kappa\approx 0.075$ a bifurcation occurs
and the extra unstable modes become stable.
Hence, in this region, it should be an attracting critical solution,
while below this range of $\kappa$, the DSS is the critical solution.
The evolutions of \cite{liebling} confirm this change in stability
and give evidence that the DSS is not the attractor above $\kappa\approx 0.1$.
The harmonic map then has both the CSS and DSS as attracting critical solutions
in distinct regions of parameter space.

A remarkable family of initial data (called {\em spiral} initial data here)
is presented which finds
the CSS critical solution in the region of parameter
space for which the DSS is the demonstrated attractor. 
In  this article, the reasons why this family can find a multiply
unstable critical solution are studied. It is argued that the
spiral initial data is quite special in that its saturation of
a charge bound disallows the growth of the extra unstable modes.

This family is presented in Section \ref{sec:results} along with the
critical solutions obtained. The family is then perturbed and the critical
solution is seen to deviate from pure CSS. These results are 
discussed in Section \ref{sec:discussion}, beginning
with a discussion of the DSS in a region where it ceases to
be an intermediate attractor. An explanation
of the reasons that the CSS can be found when not an intermediate
attractor follows. After the Conclusion, an Appendix provides the equations
of motion for the model.
I also note that the results 
described below were generated using a modified version of Choptuik's 
adaptive mesh-refinement code for massless scalar collapse~\cite{matt}.

\section{Numerical Results}
\label{sec:results}
Because in spherical symmetry the gravitational field has no
degree of freedom, initial data is completely determined by
the specification of the scalar field and its time derivative at
the initial time. Because the field $F(r,t)$ is complex, we can
decompose the field into its real and imaginary components
$\psi(r,t)$ and $\phi(r,t)$, respectively, and specify
their initial profiles $\psi(r,0)$ and $\phi(r,0)$. Setting
the generalized time derivatives (see Eq. \ref{eq:idata}) to zero initially
via
$\Pi_\psi(r,0)=\Pi_\phi(r,0)=0$ yields time symmetric data.
Setting $\Pi_\psi(r,0)=\psi(r,0)'$ and $\Pi_\phi(r,0)=\phi(r,0)'$
yields initial data which approximates an ingoing wave. For what follows,
the choice of either of these does not affect the critical solution found.

Generally the type of the initial data does not affect the obtained
critical solution because of universality, and hence a common choice
has been a Gaussian pulse in each of the components
\begin{eqnarray}
\psi(r,0) & = & A_\psi ~ e^{ - \left( r - R_\psi \right)^2 / d^2_\psi }
\nonumber\\
\phi(r,0) & = & A_\phi ~ e^{ - \left( r - R_\phi \right)^2 / d^2_\phi },
\label{eq:gaussian}
\end{eqnarray}
where $A_\psi,A_\phi,R_\psi,R_\phi,d_\psi,d_\phi$ are arbitrary real constants.
However, instead of a decomposition into real and imaginary parts, the complex
field can be expressed by a magnitude and phase 
\begin{eqnarray}
F(r,0) = f(r,0)~e^{ih(r,0)}
\label{eq:spiral}
\end{eqnarray}
for arbitrary real functions $f(r,t)$ and $h(r,t)$.

The family which is of interest here
 is most easily expressed in
this form where the phase is linear in $r$
\begin{equation}
F(r,0) = f(r) e^{i \omega r}
\end{equation}
and where $\omega$ is an arbitrary real constant.
This data represents a generalized spiral in the complex plane
and is called {\em spiral} initial data here.
So that the fields are regular and compact, $f(r)$ is constructed
such that $f(r\rightarrow 0) =0$ and $f(r_{\rm max}) =0$. The
constant $r_{\rm max}$ represents the size of the numerical grid. Also,
for reasons that should become clear later, $f(r)$ is constructed
so that it varies much more slowly in $r$ than the exponential term.
Generally, $f$ is either Gaussian or takes a step-function-like form
\begin{equation}
f(r)   =  \frac{1}{4} \Bigl[
              1 + \tanh \left( r - r_{\rm low } \right)
             \Bigr]
             \Bigl[
              1 - \tanh \left( r - r_{\rm high} \right)
             \Bigr]
\end{equation}
for arbitrary constants $r_{\rm low } < r_{\rm high}$.

\begin{figure}
\epsfxsize=7.5cm
\centerline{\epsffile{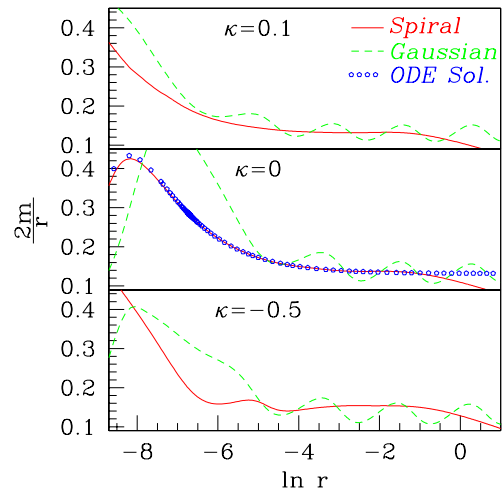}}
\caption[Critical solutions obtained for Gaussian
and spiral initial data along with the CSS of Hirschmann and Eardley.]
{Critical solutions for both
Gaussian (dashed, Eq. \ref{eq:gaussian}) and
spiral (solid, Eq. \ref{eq:spiral})
initial data for $\kappa=0.1, 0.0, 0.5$.
The solution
shown with open circles is that found by Hirschmann and Eardley with
the assumption of continuous self-similarity [5].}
\label{fig:css_to_dss}
\end{figure}

Spiral initial data is remarkable because critical searches
conducted with this data find the CSS critical solution
for $\kappa \approx 0$. 
Figure \ref{fig:css_to_dss} displays the critical solution obtained
with various initial data for values of $\kappa$ for which
the DSS is the attractor. These results show that the spiral
data is quite special in the space of initial configurations.

\begin{figure}
\epsfxsize=7.5cm
\centerline{\epsffile{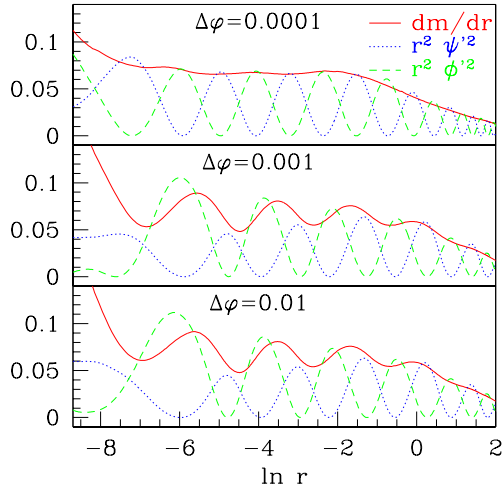}}
\caption[Perturbations of phase for the spiral initial data for $\kappa=0$.]
{Perturbations of phase for the spiral initial data for $\kappa=0$.
Even for small changes in the relative phase of $\psi$ and $\phi$, the critical
solution is driven toward the DSS.}
\label{fig:spiral_pert_p}
\end{figure}

\begin{figure}
\epsfxsize=7.5cm
\centerline{\epsffile{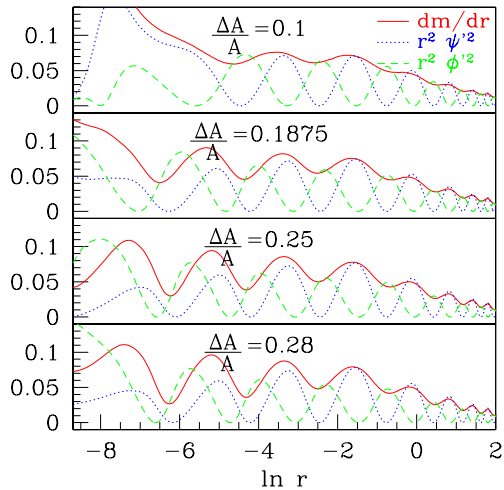}}
\caption[Perturbations of relative amplitude for the spiral initial data for $\kappa=0$.]
{Perturbations of relative amplitude for
the spiral initial data for $\kappa=0$. Again, the solution is driven
toward the DSS.}
\label{fig:spiral_pert}
\end{figure}

Perturbations of this data are made according to
\begin{eqnarray}
\psi(r,0) & = & f(r)~\left(A+\Delta A\right)~
                   \cos \left( \omega r +\Delta \varphi \right)
\nonumber \\
\phi(r,0) & = & f(r)~A~\sin \left( \omega r \right),
\label{eq:spiral_per}
\end{eqnarray}
and the critical solutions obtained are shown in Figs. \ref{fig:spiral_pert_p}
and \ref{fig:spiral_pert}.
The figures show the  last subcritical solution obtained by a critical
search at a time just before it decides to disperse. The graphs then
represent the outgoing record of the collapse of the self-similar regime
towards $\ln r \rightarrow \infty$. Large $r$ then represents early time,
and the graphs show that as the perturbation is increased, the self-similar
pulse gradually develops a discrete oscillation. This development represents
the funneling of the solution away from the CSS and toward the DSS.

These results make  clear that
changing the relative phase of the two fields or their relative amplitudes
causes the critical solution to be attracted to the DSS.
Perturbations of
the relative frequency produces similar results.

\begin{figure}
\epsfxsize=7.5cm
\centerline{\epsffile{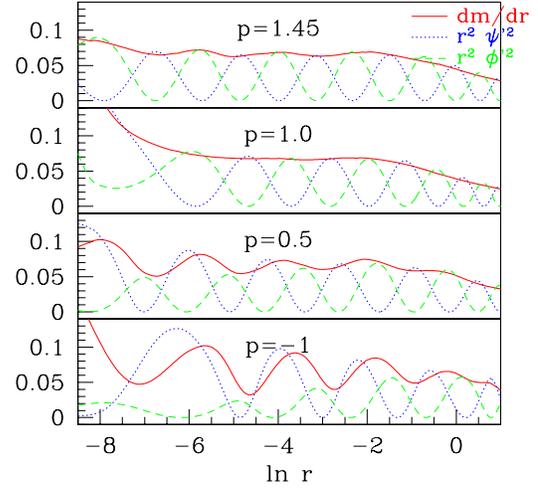}}
\caption[Critical solutions with $\kappa=0$ for spiral initial
data with exponents of $p=-1,0.5,1,1.45$.]
{Critical solutions ($\kappa=0$) 
obtained for the spiral data for various
values of $p$ [Eq. (\ref{eq:spiral_rpow})].
The initial data for these critical solutions are shown in
Figure \ref{fig:spiral_pert_power_init}.
}
\label{fig:spiral_pert_power}
\end{figure}

\begin{figure}
\epsfxsize=7.5cm
\centerline{\epsffile{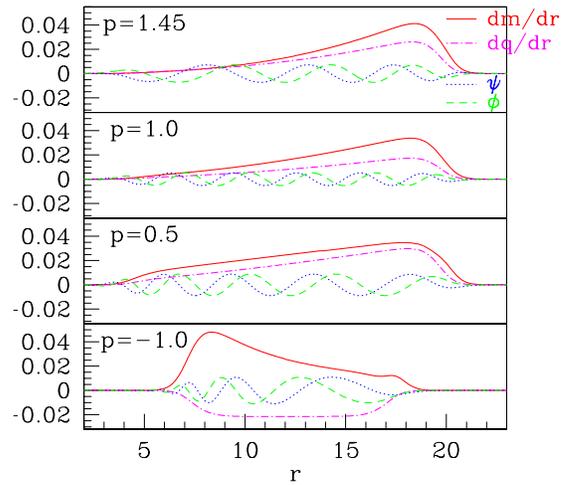}}
\caption[Initial data for the critical solutions displayed in
Figure \ref{fig:spiral_pert_power}.]
{Initial data for the critical solutions displayed in
Figure \ref{fig:spiral_pert_power}.
}
\label{fig:spiral_pert_power_init}
\end{figure}

While Figures \ref{fig:spiral_pert_p} and \ref{fig:spiral_pert} show
that
only initial data completely out
of phase finds the CSS, the spiral data can be further 
perturbed via
\begin{equation}
F(r,0) = f(r) \exp\left(i \omega r^p\right)
\label{eq:spiral_rpow}
\end{equation}
for $p\ne1$ and still be considered $\pi/2$ radians out of phase. However, as shown
in Fig. \ref{fig:spiral_pert_power}, for $p\ne 1$ the CSS is not
the critical solution. The initial data for these configurations
are shown in Fig. \ref{fig:spiral_pert_power_init}.

\begin{figure}
\epsfxsize=7.5cm
\centerline{\epsffile{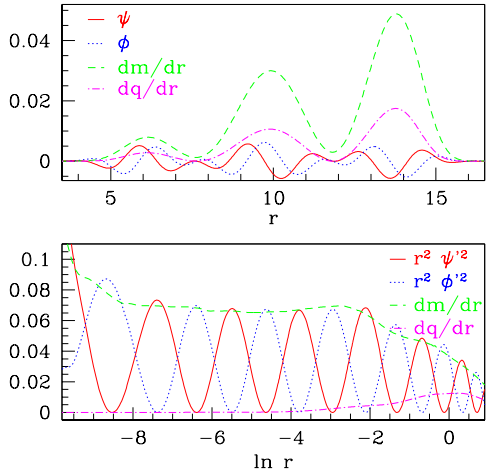}}
\caption
    {Critical solution obtained for initial data consisting
     of two frequencies $\omega_1=2.1$ and
     $\omega_2=3.6$ [Eq. (\ref{eq:two_freqs})]. The top frame shows
     the initial data.
     The bottom frame displays the
     critical solution.}
\label{fig:two_freqs}
\end{figure}

Before discussing these results, it is interesting to consider
initial data which consists of
the superposition of two different frequencies $\omega_1$ and $\omega_2$
\begin{equation}
F(r,0) = f_1(r) e^{i \omega_1 r} + f_2(r) e^{i \omega_2 r}.
\label{eq:two_freqs}
\end{equation}
Both the initial data and the critical solution are shown in
Fig. \ref{fig:two_freqs}.
This example of the superposition of two frequencies
finds the CSS as shown in the figure, however, not all
examples of this family do. Families with $f_1=f_2$
and comparable frequencies resulted in critical solutions which
appeared CSS.

As mentioned previously, initial data which is otherwise
spiral but has $f(r)$ varying on scales comparable to $\omega$
is driven away from the CSS critical solution in much the same
way that the perturbations shown in Figures \ref{fig:spiral_pert_p}
and \ref{fig:spiral_pert} drive the solution away from the CSS.
Also, many examples of  initial data of the form (\ref{eq:two_freqs})
are similarly driven away from the CSS. Experimentally, the
strongest indicator of initial data which will find the CSS is
when the energy density is everywhere proportional to the charge
density. That this proportionality indicates the specialness of
spiral initial data is discussed in Section \ref{sec:css}.

\section{Discussion}
\label{sec:discussion}

Surprisingly enough, in this case a discussion of  the DSS
critical solution in the region $\kappa \agt 0.1$ is simpler
than the discussion of the CSS.
In this region of parameter space, it is the DSS which
has multiple unstable modes, and it is relatively easy
to understand the families of initial data which find the
DSS.
Hence, the discussion of these special families
is presented first, followed by a discussion of the specialness
of the spiral data.

\subsection{The DSS}
\label{sec:dss}
In the harmonic map model, for $\kappa \agt 0.1$ the CSS
is the attracting critical solution. Numerical evolutions of
various families of initial data generically find the CSS,
and do not find the DSS. Gundlach showed that the DSS has
only one unstable mode for $\kappa=0$, but the stability
analysis of the DSS has not been done for general $\kappa$ \cite{gundlach}.
However, evolutions of this model,  as well as results in the equivalent
region of the model in \cite{liebling}, clearly indicate that for $\kappa \agt 0.1$
the DSS has more than one unstable mode.

However, there are non-generic families that do find the DSS in
this regime despite the presence of these extra unstable
modes. Consideration of this phenomenon is helpful in understanding
the CSS occurring for $\kappa=0$.

One description of initial data that finds the DSS in this regime
is mentioned in \cite{liebling}. In that work it was found that
when one component of the field was initially vanishing, it remained
zero. The model here, being equivalent to that one for $\kappa>0$, retains
this feature as shown in the equations of motion for the two components
of the scalar field in Equation (\ref{eq:waveeqs}).
Because the CSS is necessarily complex (it has charge), initial data
with one field initially vanishing is unable to find the CSS as its critical
solution.
Thus, families of initial data of the form
\begin{eqnarray}
F(r)        & = & f(r)     \nonumber         \\
\psi(r)     & = & f(r)     \label{eq:vanish} \\
\phi(r)     & = & 0        \nonumber
\end{eqnarray}
for arbitrary $f(r)$ will only find the DSS.

However, a more general set of families can be found with this
principle in mind. Consider initial data of the form
\begin{eqnarray}
F(r)        & = & e^{iC}~f(r)             \nonumber         \\
\psi(r)     & = & \cos \left( C \right) f(r)     \label{eq:inphase} \\
\phi(r)     & = & \sin \left( C \right) f(r)     \nonumber 
\end{eqnarray}
for arbitrary constant $C$. This data corresponds
to a global rotation of Eq. (\ref{eq:vanish})
by an angle $C$ in the complex plane. Because the Lagrangian
is invariant with respect to this rotation, the critical solution must
be the same as the initial data described by Equation (\ref{eq:vanish}).

A more physical understanding of this can be gained by examining the
issue of charge. For both sets of initial data (\ref{eq:vanish},\ref{eq:inphase}), the charge is zero. In fact, all components of the current density,
Equation (\ref{eq:current1}), vanish
\begin{equation}
j_\mu = 0.
\end{equation}
The divergence of this current is zero, so the current density will not
grow if initially vanishing.
In other words, the system with no charge is in a symmetric
state with respect to charge, and this symmetry would have to be broken
were the charge to become positive or negative.

Because the CSS can have
either positive or negative charge, these sets of initial data are precisely
balanced between trajectories that would take them to the CSS with
positive charge and those that would take them to the negatively charged
CSS (since the model is independent of which field is considered the
imaginary component and which the real component of the complex scalar field,
the sign of the charge of the CSS is arbitrary).
It is this balance that enables
them to {\em see} the DSS as a critical solution with only one unstable
mode when in fact it has more than one. The initial data has already
tuned one of the extra unstable modes (or an unstable conjugate pair of
modes).

With the knowledge that the extra unstable mode corresponds to a charged
mode, a two-parameter search can now be conducted. Determination of
an appropriately parameterized initial data family is somewhat more
subtle than that for the one parameter data. With a one parameter search,
a parameterization needs to be smooth and monotonic in the 
mass of the initial data near the critical point
because the excitation of the black hole mode is characterized
by the mass contained. Here, that mode must also be tuned, but the unstable
charged modes must be tuned as well. Hence, the second parameter must
be locally monotonic in charge near the critical point.

A two parameter search with the data 
\begin{eqnarray}
f(R)        & = & e^{ - \left( r - R \right)^2 / D^2 } \nonumber \\
\psi(r)     & = & p~f(R_\psi)                          \label{eq:twoparam} \\
\phi(r)     & = & p~f(R_\psi-\delta+10p)               \nonumber 
\end{eqnarray}
finds the DSS. Here, $f(R)$ is a Gaussian pulse centered on some radius $R$, and
$R_\psi$ and $D$ are arbitrary constants. The parameters $p$ and $\delta$
are used to tune both the mass and the charge of the initial data.
For $\delta$ fixed, $p$ effectively tunes the initial energy content.
With $p$ fixed and with $\delta \approx 10p$, increasing $\delta$
increases the charge from some negative value up to some positive value
(the constant $10$ is chosen arbitrarily).
Setting $\delta = 10p$ turns the family back into one with zero charge.

With this data, some value for $\delta$ near the value $10p$ is chosen,
and the black hole
critical solution is bracketed sufficiently closely so that the sign of
the initial charge of the critical solution is known.
A different $\delta$ is chosen so that
the sign of the charge of the critical solution is found to be the
opposite sign. These two values of $\delta$ then bracket a critical
value $\delta^*$ for which the critical solution ($p=p^*$) has zero charge.
In this manner, the DSS is found.

The difference between this two-parameter tuning and using initial
data with zero charge is simply the order in which the tuning occurs.
In the latter, the initial data is already tuned to have zero charge.
The remaining task is then to tune the mass. However in the former,
the charge is being tuned first to arrive at a one-parameter family
that in general has charge. It is only at the critical value of $p$
that this one-parameter family (found from the $\delta$ search)
that the solution has zero charge.

\begin{figure}
\epsfxsize=7.5cm
\centerline{\epsffile{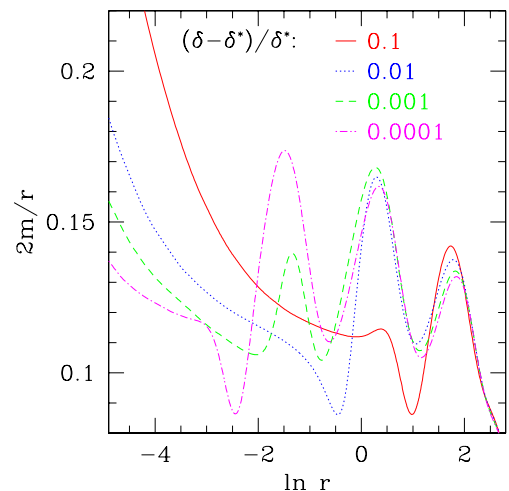}}
\caption[Critical solutions for $\kappa=1$ obtained by
tuning both charge and mass.]
{Critical solutions with $\kappa=1$
for various levels of charge tuning
$\left( \delta - \delta^* \right) / \delta^*$.
The initial data is of the form found in Eq (\ref{eq:twoparam}).
The field $2m/r$ is shown for $\left( \delta - \delta^* \right) / \delta^*$
equal to ${0.1,0.01,0.001,0.0001}$.
As $\delta$ approaches $\delta^*$ the solution appears discretely
self-similar longer but eventually becomes the CSS.}
\label{fig:twomodes}
\end{figure}

In Figure \ref{fig:twomodes} a series of critical solutions for $\kappa=1$
with initial data of the form (\ref{eq:twoparam}) are shown.
For each case, $|p-p^*|$ is of order machine precision.
The critical solution gets closer to the DSS in the limit
$\left( \delta - \delta^* \right) / \delta^* \rightarrow 0$.
Interestingly,
while the critical solutions of the perturbed spiral data
in Figure \ref{fig:spiral_pert} appeared
initially to be headed toward the CSS only to be funneled to the DSS,
these solutions do the opposite. They begin initially as DSS solutions
but eventually head to the CSS. As they get closer to the critical solution
$\delta^*$, the DSS lasts for a progressively longer time. This similarity
is consistent with the spiral being tuned to the unstable modes of the CSS.

\subsection{The CSS}
\label{sec:css}

Initial data of the form
Equation (\ref{eq:inphase}) will always  find the DSS, because,
having no charge, a direction must be picked toward either
positive or negative charge to break the symmetry.
In the case of finding the multiply-unstable CSS, the spiral data
maximizes the charge of the initial data for a given energy,
and so it appears that a symmetry must be broken to disperse all
the charge and arrive at the DSS.

To find a bound on the charge, consider the norm of the vector
$\left[F^a F^a F^c_{,\mu}
     \pm
\frac{\left(1-\kappa |F|^2\right)^2}{2} \epsilon_{cd} F^d j_\mu \right]$
at the initial time for time-symmetric initial data. Because at the
initial time all the time
components vanish, the norm is positive definite.
This is similar to a trick employed by Belavin and Polyakov \cite{belavin}
and described in Rajaraman \cite{rajaraman}.
Computing
\begin{eqnarray}
\left[F^a F^a F^c_{,\mu}   \pm \frac{\left(1-\kappa |F|^2\right)^2}{2} \epsilon_{cd} F^d j_\mu
\right]
~~~~~~~~~~~~~~~~&&\nonumber\\
\times \left[F^a F^a F^c{}^{,\mu} \pm \frac{\left(1-\kappa |F|^2\right)^2}{2} \epsilon_{ce} F^e j^\mu
\right]
& \ge & 0
\label{eq:bounda} \\
    |F|^2 F_{,\mu} F^*{}^{,\mu}
 \ge 
  \frac{\left(1-\kappa |F|^2\right)^4}{4} j_\mu j^\mu.
& &
\label{eq:bound}
\end{eqnarray}
This inequality (\ref{eq:bound}) can be expressed in terms of the energy density for
time-symmetric initial data
\begin{equation}
T_{00} = \frac{ \alpha^2 F_{,\mu} F^*{}^{,\mu} }{ 8 \pi \left(1-\kappa |F|^2\right)^2}.
\end{equation}
The bound is then
\begin{equation}
\frac{32 \pi |F|^2}{\left(1-\kappa |F|^2\right)^2} T_{00} \ge \alpha^2 j_\mu j^\mu
\end{equation}
which, for $\kappa=0$, is simply
\begin{equation}
32 \pi |F|^2 T_{00} \ge \alpha^2 j_\mu j^\mu.
\end{equation}
Hence there is an upper limit to the square of the charge density
for a given energy density. 

Letting the initial data be of the general form (\ref{eq:spiral}),
the condition to saturate this bound for $\kappa=0$ is
\begin{equation}
    |F|^2 F_{,\mu}F^*{}^{,\mu}
 = 
    \frac{1}{4} j^\mu j_\mu
\end{equation}
which implies
\begin{equation}
    \left(f'\right)^2
 =   0.
    \label{eq:spiral_restrict}
\end{equation}
Saturation therefore occurs when all energy occurs in the
phase rotation, $h'$. Physically, this is apparent
by looking at the behavior of $F$ in the complex plane.
Because $f'$ vanishes, as $F$ is traced out for various $r$,
the magnitude $f$ does not change; only the phase is changing
so the path is a circle on the plane.
This tracing then maximizes the area covered for a given
energy. The area covered is proportional to the charge, so
the charge is maximized.

This analysis applies only for $\kappa=0$ with time-symmetric
initial data, though presumably similar arguments would hold
for the generalization to other values of $\kappa$ and ingoing
initial data.

At this point, it is interesting to compare the family~(\ref{eq:inphase})
to the spiral initial data. These two families are, in some sense,
complementary. Pick some $r$, and imagine that point
in the complex plane. To construct data of the form~(\ref{eq:inphase}),
determine the field values for all other $r$ by requiring these points
to fall on the radial line between this initial point and the origin.
In this fashion, the initial data will have some global phase, constant in $r$,
equal to some value $C$. However beginning once again from
that initial point in the plane,
the restriction of Equation~(\ref{eq:spiral_restrict})
says that to construct spiral initial data, as $r$ is increased the curve
in the plane must
lie everywhere {\em perpendicular} to the radial direction.

Another condition restricting the spiral data appears to be that the
charge density must be independent of $r$. 
Just as seen with families that find the multiply unstable DSS
where the charge density must be everywhere zero independent of $r$,
here the charge density must be everywhere a maximum and independent
of $r$.

In order for the charge density to be independent
of $r$, the rate at which the data covers the plane, $\omega = h'$,
must be constant in $r$.
Initial data which would otherwise
be spiral but with $h'\ne 0$ are shown in Figure \ref{fig:spiral_pert_power_init}
with their respective critical solutions shown in
Figure \ref{fig:spiral_pert_power}. These results indicate that
only for $p=1$ in Equation (\ref{eq:spiral_rpow}) is the CSS found.

Another perspective on the constraints of the spiral data
is afforded by examining the ratio of the charge density to
the energy density
\begin{equation}
\frac{j_\mu}{ F_{,\mu} F^*{}^{,\mu} }
=
\frac{2a^2 f^2 h'}{ \left( f' \right)^2 + f^2 \left( h' \right)^2 }
=\frac{2a^2}{h'}.
\end{equation}
That this ratio is approximately independent of $r$ for the conditions
$f'=0$ and $h''=0$  appears 
to indicate that tuning the mass of the initial data also tunes
the charge. In other words, with the charge to mass ratio everywhere
the same, the critical search cannot find a solution which disperses
all the charge.

Construction of smooth, compact, and regular initial data
consistent with the restrictions 
\begin{equation}
f'(r,0)=0~~~~~~~h''(r,0) = 0
\label{eq:restrict}
\end{equation}
is quite
difficult. Regularity at the origin requires either $f(0,t)=0$
or $f'(0,t) = 0 = h'(0,t)$.

Satisfying the former along
with strict observance of~(\ref{eq:restrict}) leads to the trivial
solution $F(r,0)=0$. Instead, as mentioned earlier, initial
data is used where $f$ vanishes at the origin, but is then 
``turned on'' at some larger $r$. The condition $f'=0$ is then
not satisfied everywhere, but for $f' \ll \omega$ the CSS is
still found.

Satisfying the latter condition along with $h''=0$ at the
initial time also leads to a trivial solution $F(r,0)=C$,
for some complex constant $C$. Again, $\omega$ can be ``turned
on'' at some larger $r$, but this can only be done in a smooth
way if $h''(r,0)$ is not everywhere zero.

The difficulty in constructing non-trivial,
regular, initial data which is strictly spiral has hampered the
analysis. While initial data of the form~(\ref{eq:inphase})
has the symmetry $\phi(r) = \tan\left(C\right) \psi(r)$ which
holds at all times, it is not clear if there is such a preserved symmetry here.
The approximate symmetry $\phi(r) = \tan \left(\omega r\right) \psi(r)$ holds
at the initial time for the spiral data but
does not appear to hold throughout the evolution.
Also, the bound~(\ref{eq:spiral_restrict}) is shown only
for time-symmetric initial data but ingoing spiral data also
finds the CSS.
However, the evolutions consistently show that
initial data which has the ratio of charge density to energy
density independent of $r$ will find the CSS as its critical solution.

\section{Conclusion}
\label{sec:conclusion}
A special family of initial data is described which finds
the CSS critical solution in a region of parameter space
where generic initial data finds the DSS. The specialness of this
family is discussed. A bound on the charge density for time-symmetric
initial data is found,
which this family saturates. Because the spiral data maximizes
the charge density and because this charge density is independent
of $r$, it is argued that the extra unstable modes cannot grow
because their growth would pick a direction in which to decrease
the charge. The inability of the extra modes to grow then
indicates that the spiral data ``see'' only the black hole mode
around the CSS.

Because both the CSS and DSS can both be found where
they have multiple unstable modes, it seems possible
that other suitably tuned families in other models might
find other, previously unknown  multiply unstable critical solutions.

\section*{Acknowledgments}
\label{sec:ack}
I would like to thank Matthew Choptuik and Eric Hirschmann
for helpful discussions. I also thank Hirschmann for providing me
with his data for the CSS solution shown in Fig.~\ref{fig:css_to_dss}.
This work has been supported by NSF grants PHY 9722068
and PHY 9318152, and some computations were performed on the
facilities at the Texas Advanced Computing Center~(TACC), a member
of the National Partnership for Advanced Computational Infrastructure~(NPACI).

\appendix
\section*{Equations of Motion}
\label{sec:app}
The action $S$ for the model under study here is
\begin{equation}
S= \int d^4x \sqrt{-g} \left(
      R -\frac{2|\nabla F |^2}{\left( 1 - \kappa |F|^2 \right)^2 }
   \right),
\label{eq:action}
\end{equation}
defined in terms of a complex scalar field $F(r,t)$
\begin{equation}
F(r,t)=\psi(r,t) + i \phi(r,t),
\end{equation}
its complex conjugate $F^*(r,t)$, and a dimensionless parameter $\kappa$.
The operator $\nabla$ represents the covariant derivative.
The action~(\ref{eq:action}) is invariant with respect to global
rotations of $F$, and thus has a conserved current
\begin{equation}
j_\mu  =  \frac{i\left( F F^*{}_{,\mu} - F^* F_{,\mu} \right)}{\left(1-\kappa |F|^2\right)^2}.
\label{eq:current1}
\end{equation}
In component form where Latin indices run over $1$ and $2$ for the real and imaginary components,
this current is
\begin{equation}
j_\mu = \frac{2}{\left(1-\kappa |F|^2\right)^2} \epsilon_{ab} F^a F^b{}_{,\mu}.
\label{eq:current2}
\end{equation}

The field equations are then
\begin{eqnarray}
G_{\mu \nu} & = & 8 \pi T_{\mu \nu}
\\
\Box F & = &  \frac{-2 \kappa F^*}{1-\kappa |F|^2} F_{;\mu} F^{;\mu}
\end{eqnarray}
where $G_{\mu \nu}$ is the usual Einstein tensor and the stress
energy takes the form
\begin{equation}
T_{\mu \nu}  =  \frac{ 
       \psi_{,\mu} \psi_{,\nu} + \phi_{,\mu} \phi_{,\nu} 
     - \frac{1}{2} g_{\mu \nu} \left(  \psi_{,\rho} \psi^{,\rho}
       + \phi_{,\rho} \phi^{,\rho} \right) 
                      }{4 \pi (1-\kappa \left( \psi^2 + \phi^2 \right))^2}.
\end{equation}
In terms of the real and imaginary parts of $F$, the wave equation becomes
\begin{eqnarray}
\Box \psi & = & - \frac{2 \kappa \left[
   \psi \psi_{,\mu} \psi^{,\mu} - \psi \phi_{,\mu} \phi^{,\mu}
   + 2 \phi \psi_{,\mu} \phi^{,\mu} \right]
                       }
                   {1-\kappa \left( \psi^2 + \phi^2 \right)}
\nonumber \\
\Box \phi & = & - \frac{2 \kappa \left[
   \phi \phi_{,\mu} \phi^{,\mu} - \phi \psi_{,\mu} \psi^{,\mu}
   +  2 \psi \psi_{,\mu} \phi^{,\mu} \right]
                        }
                         {1-\kappa \left( \psi^2 + \phi^2 \right)} .
\label{eq:waveeqs}
\end{eqnarray}

We work in spherically symmetry with the metric
\begin{equation}
ds^2 = -\alpha^2(r,t)dt^2 + a^2(r,t)dr^2+r^2d\Omega^2.
\label{eq:metric}
\end{equation}
We introduce the following auxiliary variables in order to cast
the field equations in first order in time form
\begin{eqnarray}
\Pi_\psi  \equiv  \frac{a}{\alpha}\dot \psi
      & ~~~~~~~~~~~ &
\Phi_\psi  \equiv \psi'
\nonumber \\
\Pi_\phi  \equiv  \frac{a}{\alpha}\dot \phi
      & ~~~~~~~~~~~ &
\Phi_\phi  \equiv \phi',
\label{eq:idata}
\end{eqnarray}
where overdots and primes denote derivatives with respect to
$t$ and $r$, respectively.
The two second order wave equations become four, first order equations
\begin{eqnarray}
\dot \Pi_{\psi} & = &
   r^{-2}\left( \frac{r^2 \alpha}{a} \Phi_\psi \right)'
\label{evolve1}  \\
&+ &
 \frac{2 \alpha \kappa
      \left[
      \psi  \left( \Phi^2_\psi - \Pi_\psi^2
               -\Phi^2_\phi + \Pi_\phi^2 \right)
   + 2 \phi  \left( \Phi_\psi \Phi_\phi -
   \Pi_\psi \Pi_\phi \right) \right]
       }{a \left( 1-\kappa \left( \psi^2 + \phi^2 \right) \right)}
\nonumber \\
\dot \Phi_{\psi}  & = & \left( \frac{\alpha}{a} \Pi_{\psi} \right)'
\\
\dot \Pi_{\phi}  & = & r^{-2}\left( \frac{r^2 \alpha}{a} \Phi_\phi \right)'
\\
& + &
 \frac{2 \alpha \kappa
   \left[
   \phi  \left( \Phi^2_\phi - \Pi_\phi^2
             -\Phi^2_\psi + \Pi_\psi^2 \right)
   + 2 \psi  \left( \Phi_\phi \Phi_\psi -
   \Pi_\phi \Pi_\psi \right) \right]
        }
       {a \left( 1-\kappa \left( \psi^2 + \phi^2 \right) \right)}
\nonumber \\
\dot \Phi_{\phi}  & = & \left( \frac{\alpha}{a} \Pi_{\phi} \right)'.
\label{evolve4}
\end{eqnarray}
The fields $\psi$ and $\phi$ are maintained at each time step
by spatially integrating their respective spatial derivatives
\begin{eqnarray}
\psi(r,t) &  = & \int_{0}^{r} \Phi_\psi(\tilde r,t)~d\tilde r \\
\phi(r,t) &  = & \int_{0}^{r} \Phi_\phi(\tilde r,t)~d\tilde r.
\end{eqnarray}
The Hamiltonian constraint is
\begin{equation}
\frac{a'}{a}  +   \frac{a^2-1}{2r} =
   \frac{r \left[ \Phi^2_\psi + \Phi^2_\phi + \Pi_\psi^2 + \Pi_\phi^2 \right]}
        {2(1-\kappa \left( \psi^2 + \phi^2 \right))^2} .
\label{hamc}
\end{equation}
The nature of  polar slicing and the radial gauge yields the constraint
on the lapse function $\alpha$
\begin{equation}
0  =  \frac{\alpha'}{\alpha} - \frac{a'}{a} + \frac{1-a^2}{r} .
\label{slicing}
\end{equation}
Finally, combination of an evolution equation and a momentum constraint
yields an evolution equation for $a$
\begin{equation}
\dot a  =  r \alpha \frac{
        \Phi_\psi \Pi_\psi
      + \Phi_\phi \Pi_\phi
           }{ (1-\kappa \left( \psi^2 + \phi^2 \right))^2}.
\end{equation}

Regularity of the solution at the origin
demands
\begin{eqnarray}
a(0,t)' & = & \alpha(0,t)' = 0    \label{abc}\\
\Phi_\psi(0,t) & = & \Phi_\phi(0,t) = 0,
\end{eqnarray}
and local flatness there is enforced by
\begin{eqnarray}
a(0,t)=1.
\end{eqnarray}
We have the freedom to
pick a condition on $\alpha$ on each time slice which
corresponds to a global change in the labeling of slices.
We impose the condition on $\alpha$ at the large radius boundary
of the grid $r_{\rm max}$
\begin{equation}
\alpha(r_{\rm max}, t) = \frac{1}{a(r_{\rm max},t)}.
\label{alphabc}
\end{equation}
As long as no radiation is escaping from the grid,
this condition implies that coordinate time $t$ corresponds
to proper time for an observer at $r=\infty$.

The nature of being restricted to a finite grid imposes
the need for an artificial boundary condition on the matter fields there.
Since our spacetime is asymptotically flat, we impose an approximate
outgoing radiation condition on the matter fields. The flat
space wave equation for some general scalar field $\Theta(r,t)$
in spherical symmetry
\begin{equation}
\Box \Theta \Rightarrow \left(r\Theta\right)_{tt} = \left(r\Theta\right)_{rr}
\end{equation}
has the solution
\begin{equation}
r\Theta = f(r+t) + g(r-t)
\end{equation}
for two general functions
$f$ (in-going component) and $g$ (out-going component).
To eliminate the in-going component at the outer boundary, we enforce
at the boundary
the condition
\begin{equation}
\frac{ \partial \left( r\Theta \right) }{ \partial r } =
-\frac{ \partial \left( r\Theta \right) }{ \partial t }
\end{equation}
with the equations
\begin{eqnarray}
\dot \Pi_\Theta  + \Pi_\Theta'  + \frac{\Pi_\Theta}{r}  & = & 0 \\
\dot \Phi_\Theta + \Phi_\Theta' + \frac{\Phi_\Theta}{r} & = & 0 ,
\end{eqnarray}
which effectively limits reflection off the outer boundary.

The metric (\ref{eq:metric}) corresponds to a ``dynamical'' Schwarzschild
metric, allowing the association
\begin{equation}
a^2(r,t) = \left( 1 - \frac{2m(r,t)}{r} \right)^{-1},
\label{mass_co}
\end{equation}
where the field $m(\tilde r,\tilde t)$ represents the mass aspect function
measuring the amount of mass contained within a shell of
radius $\tilde r$ centered about the origin at coordinate time $\tilde t$.
Eq. (\ref{mass_co})  leads to
\begin{equation}
m(r,t)  =  \frac{r}{2} \left( 1 - \frac{1}{a^2} \right) ,
\end{equation}
so that from the field $a(r,t)$, we can obtain the mass aspect function.

An advantage of polar slicing (and most slicing conditions
used in numerical relativity) is that it avoids singularities.
Further, because of our use of polar-areal coordinates,
we cannot observe the formation of a true
horizon which is a coordinate singularity in our coordinates.
Instead, by monitoring $m(r,t)/r$ we can observe the formation
of a horizon and hence a black hole when $m/r\rightarrow 1/2$.
Where $m/r$ approaches $1/2$, we can determine the mass
of the black hole forming by simply halving the radius where
the horizon forms.


\end{document}